\documentclass[preprint,12pt]{elsarticle}

\usepackage{amssymb}

\journal{Nuclear Physics A}

\begin{document}

\begin{frontmatter}


\title{Nucleation of baryons in relativistic hadron-nucleus collisions}
 
\author{A. Ergun$^1$, N. Buyukcizmeci$^2$, A. Kittiratpattana$^{3,4}$, T. Reichert$^{5,6}$, A. S. Botvina$^{3,6}$, M. Bleicher$^{3,6,7}$}

\address{$^1$Kadinhani Faik Icil Vocational School, Selcuk University, TR-42800 Konya, T\"urkiye}
\address{$^2$Department of Physics, Selcuk University, TR-42079 Konya, T\"urkiye}
\address{$^3$Institute for Theoretical Physics, J. W. Goethe University, D-60438 Frankfurt am Main, Germany.}
\address{$^4$Center of Excellence in High Energy Physics and Astrophysics, School of Physics, Suranaree University of Technology,
University Avenue 111, Nakhon Ratchasima 30000, Thailand}
\address{$^5$Frankfurt Institute for Advanced Studies, Ruth-Moufang-Str.1, D-60438 Frankfurt am Main, Germany}
\address{$^6$Helmholtz Research Academy Hesse for FAIR (HFHF), GSI Helmholtz Center, Campus Frankfurt, Max-von-Laue-Str. 12, D-60438 Frankfurt am Main, Germany}
\address{$^7$GSI Helmholtz Center for Heavy Ion Research, Planckstr. 1, D-64291 Darmstadt, Germany}

\begin{abstract}
We suggest a new theoretical method to describe the 
baryon clusterization of nuclei in hadron-nucleus reactions. 
As an example we have explored the nuclei production 
in $\pi^-+C$ and $\pi^-+W$ collisions at 
p$_{lab}$=1.7 GeV by using the hybrid approach consisting of 
the Ultra Relativistic Quantum Dynamics Model (UrQMD) and the 
Statistical Multifragmentation Model (SMM). The UrQMD describes the 
production of new baryons, and the propagation toward the subnuclear 
densities with the fluctuations leading to the formation of excited 
baryonic clusters. The SMM describes the production of final nuclei 
and hypernuclei after interaction of baryons inside these clusters. 
We demonstrate the transverse momenta, rapidity, mass distributions 
and excitation energies of both primary clusters and final nuclei 
(including hypernuclei). The results of the UrQMD and UrQMD+SMM 
model calculations for different clusterization parameters are compared with the available HADES experimental data on
baryon production, providing a very promising window 
for future research on nuclei and hypernuclei formation in these reactions. 
\end{abstract}


\end{frontmatter}
\section{Introduction}

It is well known that in hadron-nucleus reactions 
at low and intermediate energies, between a few hundred MeV and a 
few of GeV per nucleon, many normal nuclei can be produced 
(see, e.g., a review~\cite{Bondorf:1995ua} and references in). 
As established recently in experiment and in theory, in intermediate 
energy reactions also a substantial amount of hypernuclei can be 
formed after the capture of hyperons by nuclear residues 
\cite{Bando,Arm93,Ohm97,japan,lorente,Bot16}. 
Usually the production of new nuclei is associated with the formation of 
the excited projectile-like and target-like (hyper-)residues and its following  
disintegration. Another mechanism of the nuclei formation may take 
place in relativistic nucleus-nucleus collisions, and it is related 
to the direct nucleation of the baryons which are produced during the 
dynamical collisions of particles. For normal nuclei this process is known 
since long ago \cite{Gos77}. Recent experiments at various heavy 
ion accelerator facilities have confirmed that also hypernuclei can be 
produced in such a way 
\cite{STAR:2010gyg,Rappold:2013fic,ALICE:2015oer,STAR:2017gxa,
STAR:2019wjm,ALICE:2019vlx,STAR:2021orx,Donigus:2020fon}. 
Here we will demonstrate that both 
reaction mechanisms can be understand within a hybrid approach including 
the dynamical and the statistical description. 

It is of special interest when the hypernuclei are formed around the 
threshold energy characteristic for the hyperon production. Therefore, we 
analyze the recent data obtained by the HADES experiment 
\cite{HADES:2023sre} which studied the interaction of pion beams on a 
carbon and tungsten target and allows to explore the 
production of small hypernuclear systems at relatively low beam energies. 
To this aim we have calculated the production of nuclei and hypernuclei with all 
correlations in $\pi^- +~C$ and $\pi^- +~W$ collisions at a pion 
momentum of {\bf p}$_{lab}$=1.7 GeV. To describe the first 
dynamical reaction stage 
we apply the Ultra-Relativistic Quantum Molecular Dynamics Mode (UrQMD)  
\cite{Bleicher:1999xi,Bleicher:2022kcu,Donigus:2023hiw}. To describe the 
nucleation process when the produced nuclear matter expands down to 
subnuclear densities we use the Statistical Multifragmentation Model (SMM)  
\cite{Bondorf:1995ua,Botvina:2007pd,Botvina:2020yfw,Buyukcizmeci:2020asf,Buyukcizmeci2023} 
generalized for the hypernuclear case.  
Previously, we have performed an exploratory analysis of the HADES experimental data 
in Ref.~\cite{apiwit}. In this work we focus our attention to the explanation 
of the statistical nucleation mechanism and its consequences. In particular, 
we systematically analyze the mass yields, transfer momenta, excitation 
energies of baryon clusters appearing in nuclear matter and final nuclei. 
Complementary, we demonstrate the comparison of the the UrQMD rapidity 
distributions of final free protons and $\Lambda+\Sigma^0$ with the HADES 
experimental data. The importance of the clusters and nuclei formation for all 
yields of products is discussed.

\section{Mechanism of the baryon nucleation}
 
The main description and features of the nucleation mechanism were explained 
in details in Refs.~\cite{Botvina:2020yfw,Buyukcizmeci2023,Bot22}. 
Thus we refrain from a full reiteration here and only refresh the main physical ideas 
qualitatively. The scenario is that after the collision the excited nuclear 
matter consisting of the nucleons and other species expands 
rapidly. This expansion in central ion collisions can be caused either 
by the initial dynamical compression of nuclear matter or by rescattering of 
the many free baryons produced in high energy reactions. The thermal 
expansion is also possible after the formation of excited nuclear residues in 
hadron induced reactions or in peripheral ion collisions. The 
new nuclei are then formed at low density by the local interaction of the baryons 
which are located in the vicinity in the 
phase space. Our approach includes several stages: 
1) the generation of initial distributions of produced baryons with UrQMD, 
2) identification of baryon clusters in the diluted matter, and 
3) formation of nuclei insides these excited clusters, that is described 
as the cluster's statistical decay.
In this case the hypothesis of local 
chemical equilibrium of the clusters in the expanding matter 
is very fruitful.  
The cascade-mode of the UrQMD Monte Carlo model (v.~3.5) is used, which 
means potential interactions, e.g. of Skyrme-type, are switched-off. It takes into 
account the conservation of total energy and momentum, and quantum number. 
During the UrQMD evolution of minimum bias  
$\pi^- +~C$ and $\pi^- +~W$ collisions we are looking at all nucleons 
and produced hyperons 
at the time $t=20$ fm/c after the first hadron interaction inside the 
nuclei. We have checked that using later times changes 
our final results only slightly because the interaction rate is rapidly 
decreasing. The primary hot clusters are identified 
by using the baryon relative coordinates corresponding 
to around 0.1--0.3 of normal nuclear 
density $\rho_0 \approx 0.15$ fm$^{-3}$, and baryon 
relative velocities, on the order of $v_c=0.14-0.22c$  
corresponding to the Fermi motion inside nuclei. 
For this identification we use the clusterization of baryons (CB) model 
(see Refs.~\cite{Botvina:2020yfw,Bot22,Bot15} for other details). 
In such a way the primary selected baryon clusters can be considered as 
local pieces of nuclear matter in the coexistence region of the nuclear 
liquid-gas type phase transition with a moderate temperature around 
5--8 MeV. In this region the new nuclei can be produced assuming 
chemical equilibrium. The chosen specific matter characteristics 
are crucial for the fragment formation and were obtained from a large 
body of previous studies of multifragmentation reactions in finite nuclear 
systems \cite{Bondorf:1995ua,Ogu11,EOS,Xi97,Poc97,MSU,Vio01,FASA}. 
As was demonstrated previously 
(see, e.g., Refs.~\cite{Bondorf:1995ua,Ogu11,EOS,MSU,Bot95,Bel02}) 
the statistical description of multifragmentation reactions as 
fragment production in the freeze-out volume at high energy is very 
successful, and it is 
consistent with the compound nucleus description at low energies.  
Therefore, at the last stage, the SMM 
\cite{Bondorf:1995ua,lorente,Botvina:2007pd} 
is employed to obtain final nuclei and hypernuclei.  

\section{Results of the calculations and comparison with experiment}

\subsection{Generation of initial nucleons and hyperons}

In the present model the formation of nuclei may take place only from 
nucleons and hyperons which are produced at the dynamical stage with UrQMD. 
In Fig.~1, we present the distributions of initial protons (full circles with solid lines) and 
initial $\Lambda$'s (open circles with dotted lines) after the UrQMD simulations of 
$\pi^{-} +~C$ (left panels) and $\pi^{-} +~W$ (right panels) 
collisions at {\bf p}$_{lab}$=1.7 GeV. 
The obvious size effect of the initial collision system gives higher yields 
for $\pi^- +~W$ than for $\pi^- +~C$ reaction. Another known  
effect is that the number of protons and the $\Lambda$'s decrease with increasing 
transverse momentum for both systems.
We can naturally describe the formation of baryonic clusters with smaller 
masses and high energy by considering baryons of high rapidity and assuming 
their clusterization. However, it is seen 
that in the both cases we have the maximal  proton densities in the region of the 
target rapidities. These protons are mostly spectator protons and they are 
weakly involved in the dynamical interactions. Spectator protons and neutrons may capture hyperons to form massive
excited baryonic clusters, which are usually called target
residues. These (hyper-)residues will subsequently decay into
multiple light hyper- and normal nuclei. Within our approach we include both the clusterization and 
the target disintegration process, therefore, we can describe the spectator 
fragmentation in addition to the nuclei formation from fast baryons. 
\begin{figure}[tbh] 
\begin{center}
\includegraphics[width=8.5cm]{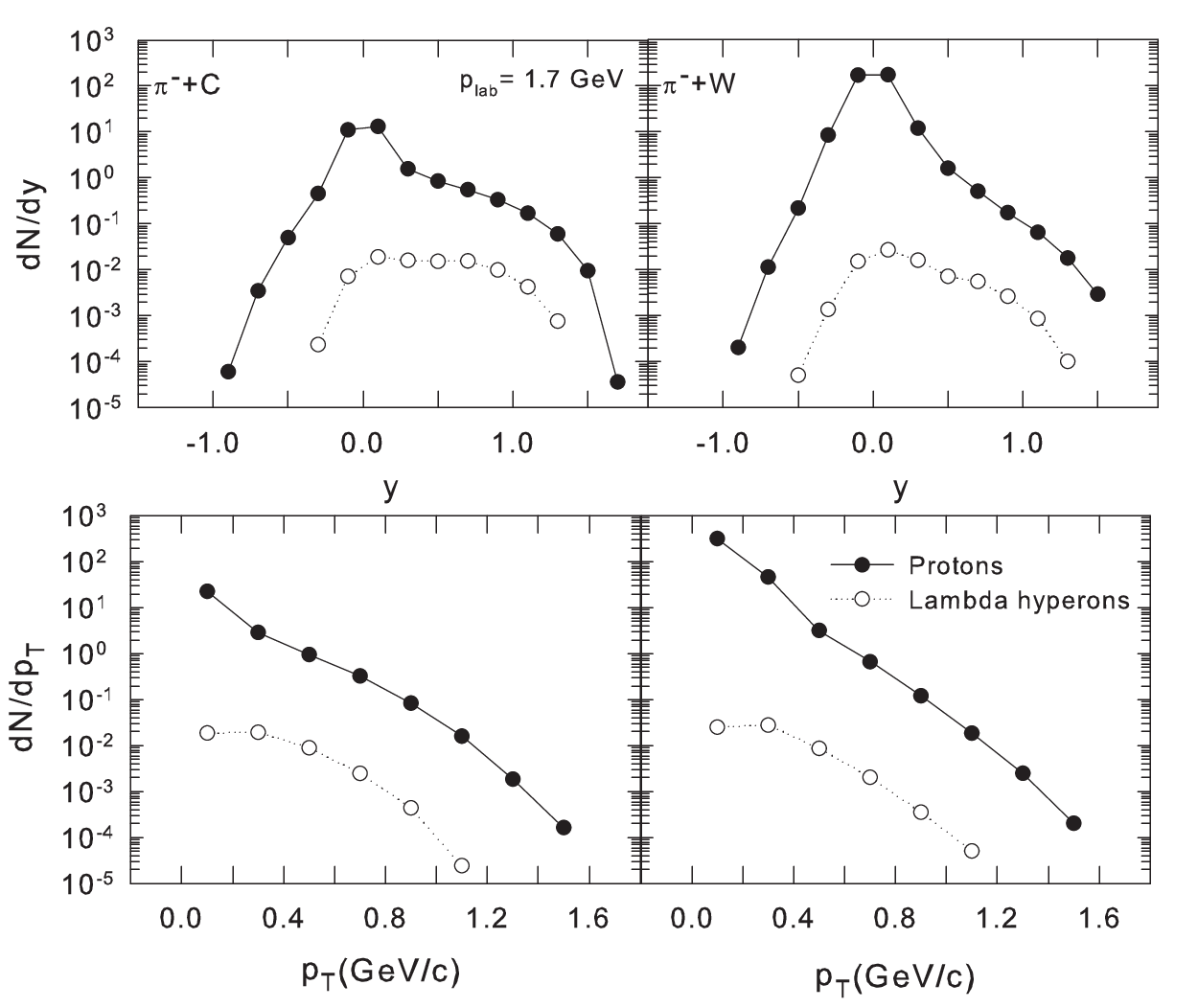}
\end{center}
\caption[]{\small{Total proton (full circles with solid lines) and $\Lambda$ (open circles with dotted lines) 
distributions (per event) after UrQMD calculations of 
$\pi^{-} + C$ (left panels) and $\pi^{-} + W$ (right panels) collisions 
at {\bf p}$_{lab}$=1.7 GeV. Top panels: rapidity distributions, bottom panels: 
transverse momentum distributions, in the rapidity range $|y| < 0.5$.}}
\label{fig1}
\end{figure}

\subsection{Mass and excitation energy distributions of baryonic clusters}

The following production of nuclei depends on the masses and excitation 
energies of the primary baryonic clusters which can be separated in the expanding 
nuclear matter because of the baryon interactions and stochastic fluctuations 
during the dynamical UrQMD stage. We have mentioned that we have used 
the CB model procedure for the cluster recognition from the produced 
baryons. Such a formation of clusters is illustrated in Fig.~2 with the 
clusterization parameters $v_c=0.22$c and $v_c=0.14$c, in the reaction 
of $\pi^- +~C$ (left panels) and $\pi^- +~W$ (right panels) 
collisions at {\bf p}$_{lab}$=1.7 GeV. The top panels show the 
mass number yield per event of hot clusters, while the bottom panels 
show their average excitation energies. The clusters are selected when the 
baryon density in the clusters has reached 1/6$\rho_0$ during the expansion 
of the matter. For futher details we refer the reader to 
Refs.~\cite{Botvina:2020yfw,Bot22}. Generally, in all cases 
there is a decrease of the yields with the mass, as in any 
coalescence-like procedure. However, when the target is small, the 
target residue may overlap with the formed clusters. 
One can see that the parameter $v_c=0.22$c gives a higher yields of 
large clusters before de-excitation than $v_c=0.14$c. However, for 
$v_c=0.22$c the excitation energies of the clusters are also higher. 
As a result, after the de-excitation the final inclusive yields of nuclei may 
be similar for both parameters. The difference between the parameters 
can still be seen in different correlations of the produced particles 
\cite{Botvina:2020yfw}. 
An adequate selection of the clusters should lead to typical excitation 
energies around 4--10 MeV per nucleon. Because, this 
range of the excitation energies, with densities around 
$0.1-0.3\rho_0$ corresponds to the coexistence region of the nuclear 
liquid-gas type phase transition. It was established in numerous 
multifragmentation studies (see, e.g., 
\cite{Bondorf:1995ua,Ogu11,EOS,Xi97,Poc97,MSU,Vio01,FASA,Bot95} 
and references in) that at these nuclear matter parameters the 
interaction between nucleons leads to the extensive production of 
nuclei. At higher densities the new nuclei can not be effectively 
singled out from nuclear matter. 
The main difference between the previous statistical description 
of multifragmentation and the present approach is that we consider 
not only one big single equilibrated nuclear source, like a residual nucleus, 
but many small clusters in local 
chemical equlibrium. In the following we demonstrate qualitatively 
how this mechanism works for the considered pion induced reactions. 
\begin{figure}[tbh] 
\begin{center}
\includegraphics[width=8.5cm]{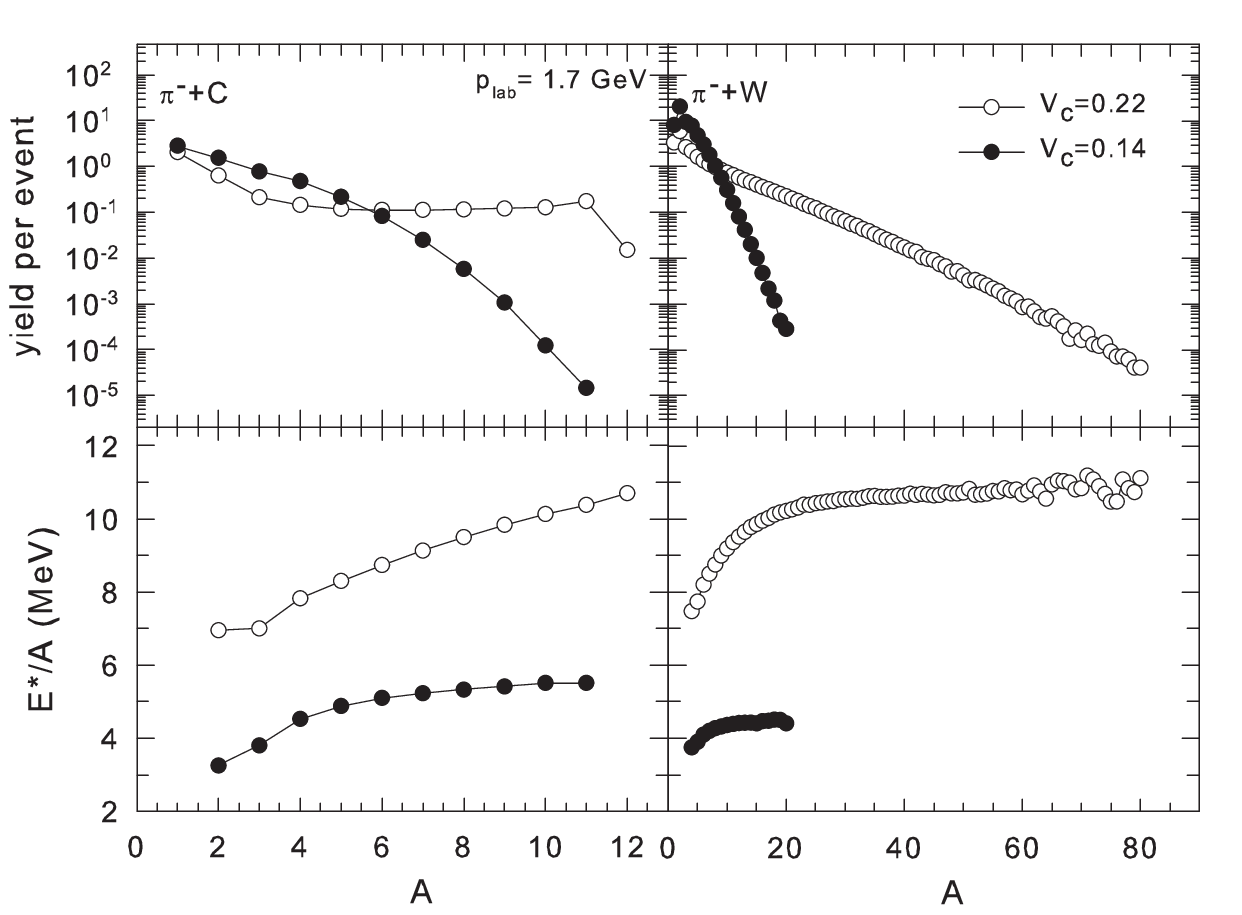}
\end{center}
\caption[]{\small{Top panels show mass distributions of the hot clusters with the 
velocity parameter $v_c=0.14$c and 0.22c after the UrQMD calculations and 
CB procedure. Bottom panels show average excitation energy of the hot clusters 
versus their mass number. Left panels are for $\pi^- +~C$ reactions and right panels are for 
$\pi^- +~W$ reactions.}}
\label{fig2}
\end{figure}

\subsection{Evolution from hot local clusters to final nuclei}

The HADES experiment can measure light nuclei. Therefore, we concentrate 
our present investigation on the predictions for light nuclei. In our approach 
light nuclei come mostly from the de-excitation of moderate-size clusters, which 
are described within SMM using the Fermi-break-up model \cite{lorente}. 
In Fig.~3 we present the normalized transverse momentum distributions of 
primary clusters (i.e., before de-excitation) with a nucleon content corresponding 
to protons, deuterons and tritons at $|y| < 0.5$. We further show the 
distributions of the same nuclei after the nucleation processes. 
Their yields are summed over all local clusters produced 
in the reaction events. One can see that the distributions of final nuclei 
are steeper and their maxima are shifted towards low momenta. This is 
a consequence of the decay of the heavy primary clusters which are 
formed by clustering from slow baryons. It is instructive that this 
effect is important even for protons. Therefore, the measured proton 
spectra at low momenta may be considerably influenced by the local interaction 
during the nuclei formation inside the clusters consisting of matter at low density. 
Let us stress that this process is beyond the initial dynamical evolution. 
\begin{figure}[tbh] 
\begin{center}
\includegraphics[width=8.5cm]{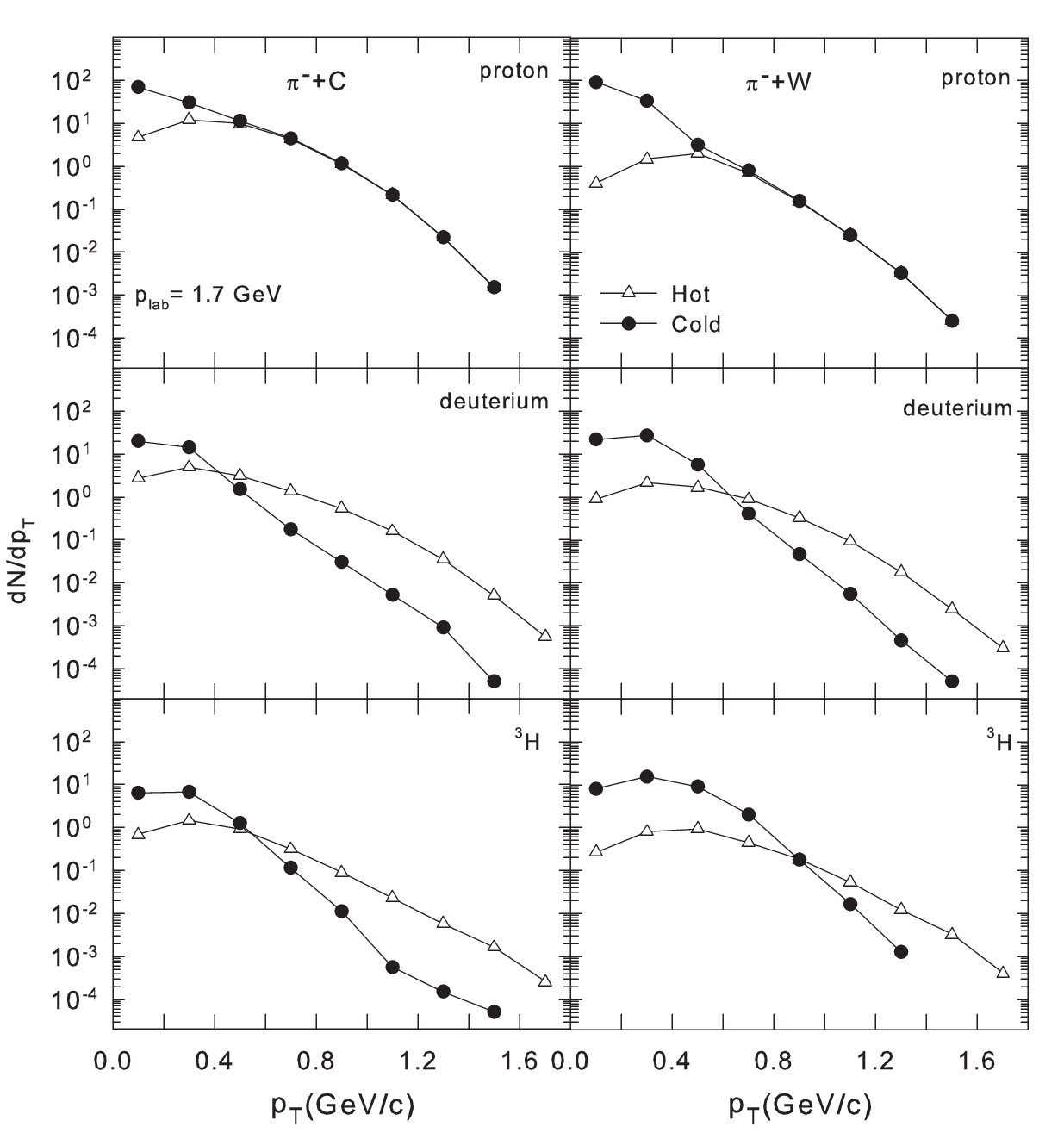}
\end{center}
\caption[]{\small{Transverse momentum distributions of primary p, d, $^{3}$H 
nuclear species obtained after subdivision of the expanding matter 
into hot primary clusters (noted as "Hot") and the same final nuclei 
(noted as "Cold"). The incident pion beam momentum is 1.7 GeV/c. The 
clusters are obtained within UrQMD+CB procedure,
and the nuclei after adding SMM de-excitation of all clusters. 
Left panels are for pion collisions with carbon, and right ones with tungsten.}}
\label{fig3}
\end{figure}

\subsection{Comparison of theory with HADES experiment and predictions}

By taking into account the dynamical reaction stage we are able to compare 
our calculations with HADES experimental data \cite{HADES:2023sre}. 
In Fig.~4 we show the comparison of the integrated proton production 
versus the rapidity, which is obtained in the experiment via the interpolation 
procedure typically adopted for the midrapidity region in relativistic heavy-ion 
collisions. In the calculation we show the UrQMD distributions for 
protons which are not involved in the following cluster formation, i.e., they 
escape from the nuclear system free. A slight enhancement in the calculations 
for protons in the target region $y \approx 0$ may be related to the detection 
cut-off in the transverse momenta and the interpolation procedure. This 
requires an additional investigation including the correlation measurements. 
It is important that the corresponding fraction of protons is only around 10 
percent in the carbon case, and even lower for the heavy target in the region of 
$|y| < 0.3$. The main fraction goes into the cluster formation. 
Generally, the comparison is encouraging, and this gives us a confidence to 
continue with the predictions for complex nuclei. 
\begin{figure}[tbh] 
\begin{center}
\includegraphics[width=10cm]{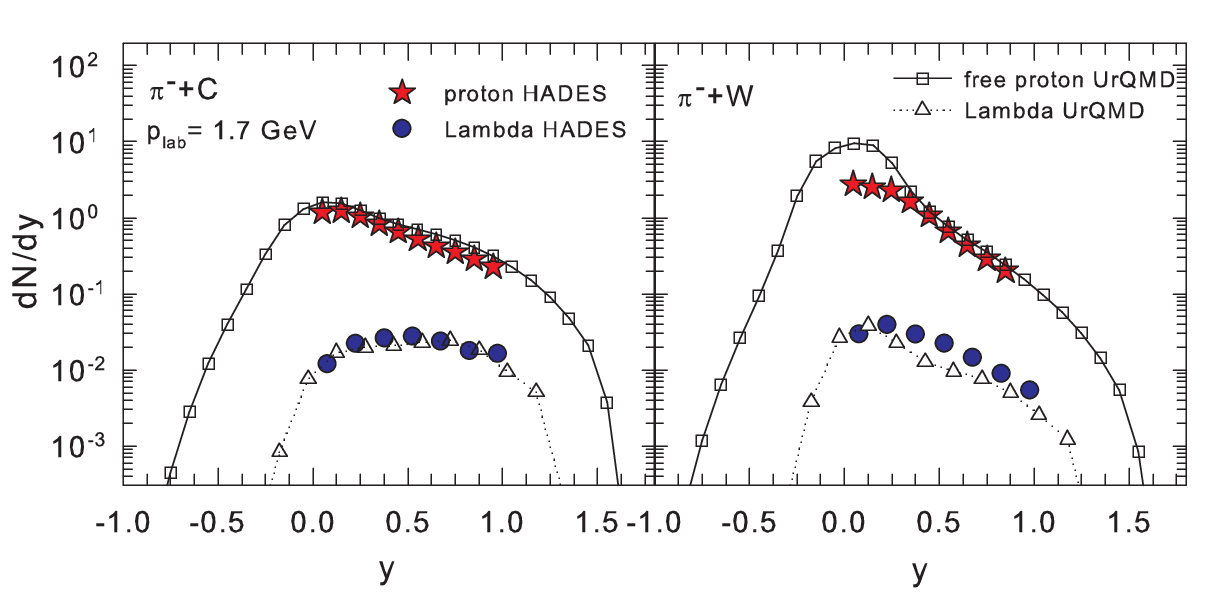}
\end{center}
\caption[]{\small{The distributions 
of protons and $\Lambda$s as a function of the rapidity for 
$\pi^- +~C$ (left panel) and $\pi^- +~W$ (right panel) collisions. 
The calculation results after UrQMD are shown for  free protons (open squares with solid lines) and $\Lambda$s
(triangles with dotted lines). Red stars and blue circles depict the HADES measurements 
\cite{HADES:2023sre}. Experimental values are divided by the total cross section 196.35 mb ($\pi^- +~C$) and 1327.32 mb ($\pi^- +~W$) used for normalization.}}
\label{fig4}
\end{figure}
\begin{figure}[tbh] 
\begin{center}
\includegraphics[width=12cm]{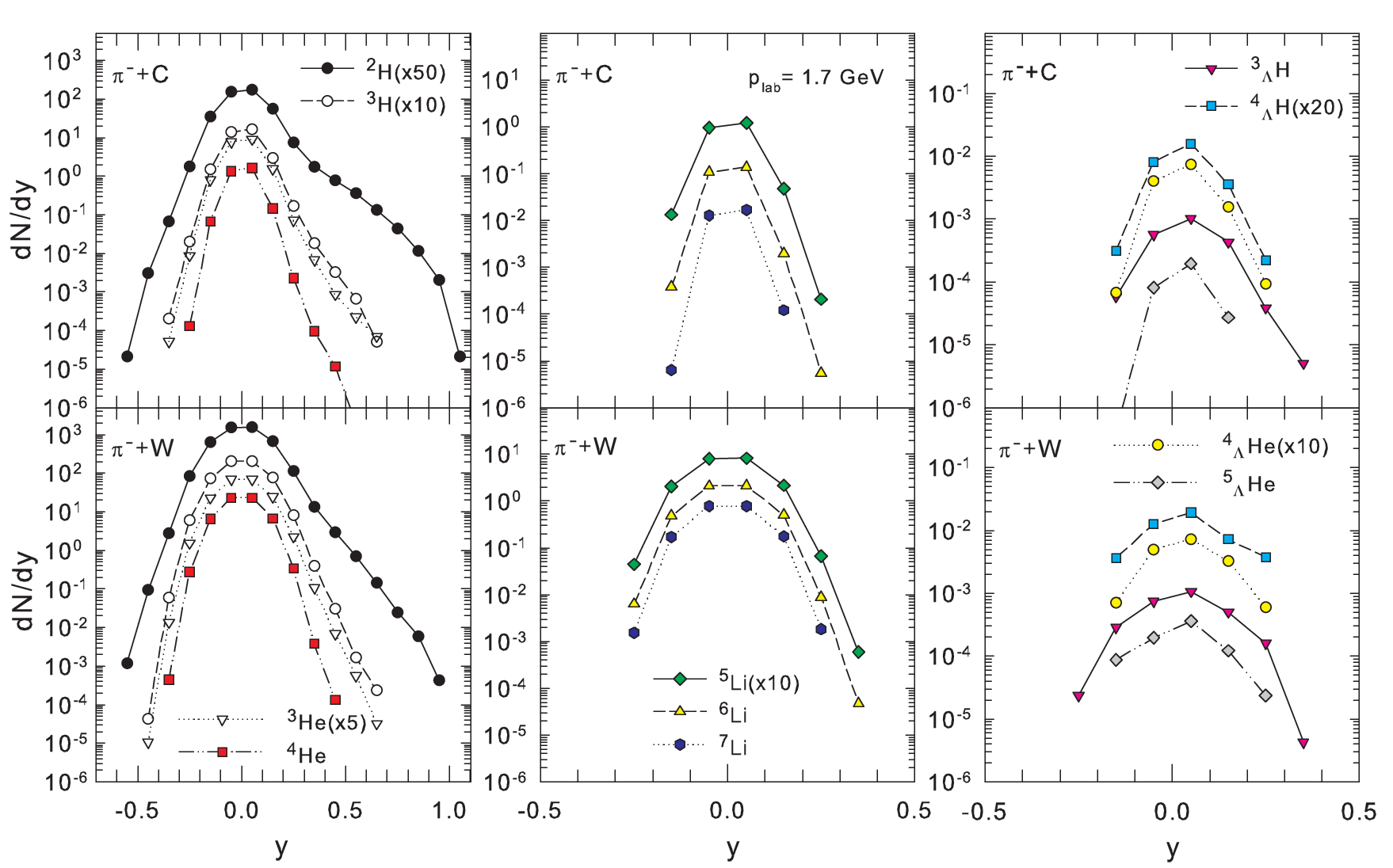}
\end{center}
\caption[]{\small{The rapidity distribution of of light nuclei 
(left and middle panels) and hypernuclei (right panels) produced in 
$\pi^- +~C$ (top panels) and $\pi^- +~W$ (bottom panels) collisions. 
The calculation results after UrQMD, CB and SMM are shown as 
coloured lines with symbols.}}
\label{fig5}
\end{figure}

In Fig.~5 we depict the rapidity distributions of the final yields 
of small nuclei and hypernuclei after performing the full de-excitation chain. 
We encourage the HADES experiment to measure yields of lightest nuclei 
(with mass number up to A=3--4). Such a measurement would be instructive to 
demonstrate that 
such a complicated many-body process of the fragment formation in pion induced 
reactions is at work. 
Previously, the phenomenological coalescence-like models and decay models for 
compound processes have been used for this purpose. 
The next step would be to measure the exclusive production of larger nuclei and 
hypernuclei (with A$>$4) with correlated particles on an event-by-event basis. 
This will provide a crucial test for the presented theoretical approach for small 
systems. Our predictions of nuclear isotopes 
including Li, and known hypernuclei including ones with A=5 are shown in Fig.~5. 
One observes rather large yields of these nuclei. Their rapidity distributions
indicate that 
this formation takes place mainly around the target kinematic region. Therefore, 
a nearly 4$\pi$ exclusive registration of the reaction products with all 
particle correlations will be possible with modern detectors. As was 
demonstrated previously \cite{Buyukcizmeci2023} the description of the yields 
of both normal nuclei and hypernuclei in correlation with other particles 
provides detailed information on the properties of normal nuclear matter and hyper-matter 
at subnuclear densities, and explain the differences between hyperon and 
nucleon interactions. 

\section{Conclusion}

We have investigated the nucleation process in nuclear matter formed 
after hadron-nucleus collisions at moderate collision energy. As an example we 
have considered the light nuclei production in pion induced reactions 
($\pi^- +~C$ and $\pi^- +~W$). We demonstrated that it is possible to 
measure nuclei produced in such reactions in the HADES and other experiments. 
A special interest of such reactions is that the formation process takes place 
in the target kinematic region, in contrast to the midrapidity region characteristic 
for central collisions of 
relativistic heavy ions. This provides a consistent connection between 
the formation of light nuclei and the well-known formation of big 
fragments which come from the multifragment disintegration of large 
excited nuclear residues. 
It was suggested that the decomposition of the excited nuclear 
matter into nuclei and hypernuclei in the collisions can be 
naturally explained by their statistical formation at subnuclear 
density when the matter expands and enters the coexistence region of 
the nuclear liquid-gas phase transition. We used 
UrQMD transport model to produce the initial baryon distributions, 
then we determined the  
baryons which participate in the excited clusters at local equilibrium, 
and then the statistical SMM was used (for the de-excitation) to describe 
the final formation of nuclei and hypernuclei inside these clusters. 

As was shown previously such a 
mechanism explain the data well in heavy-ion collisions, therefore, it may 
be extended for the hadron induced fragmentation reactions, too. We 
consider it as a universal mechanism for the nuclear fragment production. 
It is important that in this process the parameters of the 
hot nuclear matter in the clusters should coincide with the parameters 
of finite nuclear systems extracted from multifragmentation studies. 
The secondary baryon interaction and the nucleation in the diluted 
matter may also change the dynamical characteristics of free baryons 
detected in experiments. 
The comparison with the first HADES data is encouraging and it 
gives confidence that future research on the production of nuclei 
and hypernuclei in these reactions can reveal the properties of 
(hyper-)matter at subnuclear densities. 

\section*{Acknowledgements}

The authors acknowledges German 
Academic Exchange Service (DAAD) support from a PPP exchange grant 
and the Scientific and Technological Research Council of 
T\"urkiye (TUBITAK) support under Project No. 121N420. 
T.R. acknowledges support through the Main-Campus-Doctus 
fellowship provided by the Stiftung Polytechnische Gesellschaft 
Frankfurt am Main and further thanks the Samson AG for their 
support. Computational resources were provided by the 
Center for Scientific Computing (CSC) of the Goethe University 
and the "Green Cube" at GSI, Darmstadt. 
A.E. and N.B. thanks J.W. Goethe University Frankfurt am Main 
for hospitality. N.B. acknowledges that the work has been performed and supported
in the framework of COST Action CA22113 (THEORY-CHALLENGES).

\end{document}